\documentclass[a4paper,12pt]{article}
\usepackage{amssymb,amsmath,verbatim,graphics,graphicx,color,slashed,bm}
\usepackage[utf8]{inputenc}
\usepackage[normalem]{ulem}

\usepackage[sort&compress,numbers, merge]{natbib}

\definecolor{Orange}{cmyk}{0,0.61,0.87,0}
\definecolor{JungleGreen}{cmyk}{0.99,0,0.52,0}
\definecolor{OliveGreen}{cmyk}{0.64,0,0.95,0.40}
\definecolor{Brown}{cmyk}{0,0.81,1,0.60}
\definecolor{RoyalBlue}{cmyk}{0.71,0.53,0,0.12}

\usepackage[colorlinks=true, linkcolor=red,
citecolor=blue,urlcolor=blue]{hyperref} 

\graphicspath{{figures/}}

\newcommand{\GeV}{\mathrm{GeV}}

\newcommand{\eff}{\textrm{eff}}
\newcommand{\planck}{\texttt{Planck }}

\begin{document}

\begin{titlepage}

\begin{flushright}
{\tt 
KIAS-P17043 \\
UT-17-23
}
\end{flushright}
	
\vskip 1.35cm
\begin{center}

{\large \bf 
Hidden Charged Dark Matter and Chiral Dark Radiation
}

\vskip 1.5cm

P. Ko$^a$,
Natsumi Nagata$^b$,
and
Yong Tang$^b$

\vskip 0.8cm

{\it $^a$School of Physics, Korea Institute for Advanced Study, Seoul
 02455, South Korea} \\[3pt]
{\it $^b$Department of Physics, University of Tokyo, Bunkyo-ku, Tokyo
 113--0033, Japan} 

\vskip 0.5cm

(\today)

\begin{abstract}
{In the light of recent possible tensions in the Hubble constant $H_0$
 and the structure growth rate $\sigma_8$ between the \planck and other
 measurements, we investigate a hidden-charged dark matter (DM) model
 where DM interacts with hidden chiral fermions, which are
 charged under the hidden SU($N$) and U(1) gauge interactions. The
 symmetries in this model assure these fermions to be massless. The DM in
 this model, which is a Dirac fermion and singlet under the hidden
 SU($N$), is also assumed to be charged under the U(1) gauge symmetry,
 through which it can interact with the chiral fermions. Below the
 confinement scale of SU($N$), the hidden quark condensate spontaneously
 breaks the U(1) gauge symmetry such that there remains a discrete
 symmetry, which accounts for the stability of DM. This condensate also
 breaks a flavor symmetry in this model and Nambu--Goldstone bosons
 associated with this flavor symmetry appear below the confinement
 scale. The hidden U(1) gauge boson and hidden quarks/Nambu--Goldstone
 bosons are components of dark radiation (DR) above/below the
 confinement scale. These light fields increase the effective number of
 neutrinos by $\delta N_{\eff}\simeq 0.59$  above the confinement scale
 for $N=2$, resolving the tension in the measurements of the
 Hubble constant by \planck and Hubble Space Telescope if the
 confinement scale is $\lesssim 1$~eV. DM and DR continuously scatter
 with each other via the hidden U(1) gauge interaction, which suppresses
 the matter power spectrum and results in a smaller structure growth
 rate. The DM sector couples to the Standard Model sector through the
 exchange of a real singlet scalar mixing with the Higgs boson, which
 makes it possible to probe our model in DM direct detection
 experiments. Variants of this model are also discussed, which may offer
 alternative ways to investigate this scenario. 
}
\end{abstract}

\end{center}
\end{titlepage}

\section{Introduction}

Cold Dark Matter (CDM) has been one of the main paradigms to account for
the missing mass in our Universe. It provides a consistent theoretical
framework and viable explanations for the compelling patterns observed in
cosmic microwave background (CMB), large scale structure (LSS), galactic
rotation curves, and so on. On top of its success, various microscopic
models of CDM have been proposed, most of which modify the ultraviolet
behavior of the Standard Model (SM) with new, weakly-interacting
degrees of freedom. In these models, physics after Big-Bang
Nucleosynthesis (BBN) essentially does not change from the standard
cosmology since CDM decoupled earlier, except rare late-time
annihilation and/or possible decay of DM. 

The framework of CDM, however, does not fully determine particle
contents and interactions in DM models, which leave a plenty of freedom
for model building. For instance, we may consider a CDM model which
contains extra stable particles besides DM and/or some interactions that
are less relevant to the thermal relic abundance of DM particles. In
this paper, we discuss a scenario where DM interacts with other very
light particles even after the BBN time. These light particles behave as
dark radiation (DR) in the Universe. The motivation for such a scenario
is twofold: theoretically and observationally. On the theory side, DM-DR
interactions are actually found in various models, such as hidden
charged DM \cite{Feng:2009mn, Carroll:mha, Baek:2013qwa, Baek:2013dwa,
Ko:2014nha, Ko:2014bka, Choi:2015bya, Foot:2004wz, Foot:2016wvj,
Tulin:2013teo, Fan:2013yva, Agrawal:2016quu}, atomic DM \cite{Kaplan:2009de, Cline:2012is,
CyrRacine:2012fz}, composite DM \cite{Hur:2007uz, Hur:2011sv,
Hochberg:2014kqa, Cline:2013zca, Soni:2016gzf, Forestell:2016qhc}, and
so on. Our model provides a simple example for such models, which may be
embedded into a more fundamental theoretical framework. 

On the observation side, such a scenario could help to resolve some
controversies in the CDM paradigm \cite{Weinberg:2013aya,
Tulin:2017ara}; for example, 
some recent models \cite{Buen-Abad:2015ova, Lesgourgues:2015wza,
Ko:2016uft, Ko:2016fcd, Tang:2016mot} may relax the tensions in the
Hubble constant $H_0$ and the structure growth rate $\sigma_8$ obtained
in the \planck and other low red-shift measurements. The latest Hubble
Space Telescope (HST) data \cite{Riess:2016jrr} gives $H_0=73.24\pm
1.74$~km s$^{-1}$Mpc$^{-1}$, 
which is about 3$\sigma$ larger than the \planck value
\cite{Aghanim:2016yuo}. Weak-lensing surveys, such as CFHTLenS
\cite{Heymans:2012gg}, measured $\sigma_8(\Omega_m/0.27)^{0.46}=0.774\pm
0.040$ while the \planck data \cite{Planck:2015xua} yields $\sigma
_8=0.815\pm 0.009$. A more recent result on $\sigma_8$ from {\tt
KiDS-450}~\cite{Kohlinger:2017sxk} also indicates $3.2\sigma$ deviation
from the \planck value. The tension in $H_0$ can easily be relaxed if we add
some amount of radiation component with the effective number of
neutrinos of $\delta N_\eff\simeq 0.4$--1.0~\cite{Riess:2016jrr}, which
increases the CMB value of $H_0$. However, due to a positive
correlation, a larger $H_0$ tends to 
result in a larger $\sigma_8$, which makes the tension in $\sigma_8$
even worse. While extended cosmological models with more parameters
\cite{Pourtsidou:2016ico, DiValentino:2016hlg, Qing-Guo:2016ykt,
Archidiacono:2016kkh, Wyman:2013lza, Zhang:2014lfa, DiValentino:2016ucb,
Barenboim:2016lxv, Sebastiani:2016ras, DiValentino:2017iww,
Krall:2017xcw} may be able to accommodate 
these tensions, specific solutions from particle physics have also been
proposed recently in Refs.~\cite{Chudaykin:2016yfk, Anchordoqui:2015lqa,
Enqvist:2015ara, Hamaguchi:2017ihw} for decaying DM and in
Refs.~\cite{Buen-Abad:2015ova, Lesgourgues:2015wza, Ko:2016uft,
Ko:2016fcd, Chacko:2016kgg, Prilepina:2016rlq} for interacting DM and DR
where either gauge bosons or fermions are the DR so that their lightness
is protected by gauge symmetry or chiral symmetry. In the scenario of
interacting DM and DR, the scattering between DM and DR can induce
diffusion damping on the matter power spectrum of DM \cite{Boehm:2004th,
Bertschinger:2006nq, Loeb:2005pm, Bringmann:2006mu, Kamada:2016qjo,
Kamada:2017oxi}, possibly resulting in a suppressed structure growth
rate, or smaller $\sigma_8$.

In this paper, we propose a new interacting DM-DR model where hidden
SU($N$)-charged quarks constitute DR and interact with Dirac fermion
DM through a hidden U(1) gauge interaction. The symmetries in this model
forbid the mass terms for hidden quarks and thus make them massless to
be DR. Moreover, when the SU($N$) interaction becomes strong and gives rise 
to confinement, the hidden quark would condense and spontaneously break
the associated flavor symmetry, which leads to Nambu--Goldstone bosons
below the confinement scale. The hidden U(1) gauge symmetry is assumed
to be rather weak so that the flavor symmetry is a good symmetry, and
thus the resultant Nambu--Goldstone bosons are naturally light and can
behave as DR in the early Universe. The hidden quark condensate also
breaks the hidden U(1) gauge symmetry into a $\mathbb{Z}_2$ symmetry,
which stabilizes the DM in our model. When the confinement scale of the
hidden SU($N$) gauge interaction is very low, the hidden U(1) gauge
boson would be extremely light and also comprises a part of DR. The
light fields in this model contribute to the effective number of
neutrinos by $\delta N_{\rm eff} \simeq 0.59$ above the confinement
scale for $N=2$ and resolve the discrepancy in the measurements of the
Hubble constant if the confinement scale is $\lesssim 1$~eV. Moreover,
the DM-DR interactions induced by the exchange of the hidden U(1) gauge
boson suppress the matter power spectrum for wave-number $k \gtrsim
0.01$~$h/{\rm Mpc}$, and make the $\sigma_8$ measurements consistent
with each other. The DM sector couples to the SM sector through the
exchange of a real singlet scalar boson that mixes with the SM Higgs
boson, which enables us to probe this model in DM direct detection
experiments.

This paper is organized as follows. In Sec.~\ref{sec:model}, we explain
our model setup in detail. Then, in Sec.~\ref{sec:pheno}, we discuss
thermal history of this model, with estimating the DM relic density. In
Sec.~\ref{sec:drpheno}, we evaluate the abundance of DR, and discuss the
diffusion damping on the matter power spectrum of DM induced by the
DM-DR interactions. Section~\ref{sec:conc} is devoted to conclusion and
discussions.

\section{Model}
\label{sec:model}

\subsection{Lagrangian}

\begin{table}[t]
 \begin{center}
\caption{The quantum numbers of the hidden sector fields. }
\label{tab:charge}
\vspace{5pt}
\begin{tabular}{c|ccccccc}
\hline
\hline
& ~$S$~ & ~$\chi_L$~ & ~$\overline{\chi}_R$~ & ~$\Psi_1$~ & ~$\Psi_2$~ 
& ~$\overline{\Psi}_1$~ & ~$\overline{\Psi}_2$~ \\
\hline
~SU($N$)~ & ${\bf 1}$ & {\bf 1} & {\bf 1} & ${\bf N}$ & ${\bf N}$ 
& $\overline{\bf N}$ & $\overline{\bf N}$ \\
U(1) & 0 & $+1$ & $-1$ & $Q_\Psi$ & $-Q_\Psi$ & $-(Q_\Psi -2)$ & $Q_\Psi
			     -2$ \\
U(1)$_B$ &0 & $0$ & $0$ & $+1$ & $+1$ & $-1$ & $-1$ \\
\hline
\hline
\end{tabular}
 \end{center}
\end{table}

To begin with, let us present the model considered in the following
discussion. We introduce a real singlet scalar $S$, a Dirac fermion
$\chi$, and four Weyl fermions $\Psi_1$, $\Psi_2$, $\overline{\Psi}_1$, and
$\overline{\Psi}_2$. These additional fields are singlets under the SM
gauge symmetry. Besides the SM gauge symmetry, this model has the
hidden $\text{SU}(N) \otimes \text{U}(1)$ gauge symmetry, under which
all of the SM fields are singlets. The singlet scalar field $S$ has no
charge under both the SM and hidden gauge symmetries. The Dirac
fermion $\chi$, which is regarded as DM in our model, is also singlet
under the $\text{SU}(N)$, but has the U(1) charge $+1$. $\Psi_1$ and
$\Psi_2$ are fundamental representations of $\text{SU}(N)$ with the U(1)
charge $Q_\Psi$ and $-Q_\Psi$, respectively. We here assume $Q_\Psi \neq
1$. $\overline{\Psi}_i$ ($i =1,2$) are anti-fundamental representations
of SU($N$) and have the U(1) charge $-(Q_\Psi -2)$ and $Q_\Psi -2$,
respectively. Thus, this model is a chiral gauge theory for $Q_\Psi \neq
1$ and $N>2$. One can easily demonstrate that this model is free from gauge
anomaly. The quantum numbers of the extra fields are summarized in
Table.~\ref{tab:charge}. Here, the fermion fields are described in terms
of left-handed Weyl fermions; in particular, the Dirac DM field is
decomposed as $\chi = (\chi_L, \overline{\chi}_R^\dagger)$. We also
show the assignment of the hidden baryon number U(1)$_B$, which is a
global U(1) symmetry in the dark sector---$\Psi_i$ and
$\overline{\Psi}_i$ have the hidden baryon number $+1$ and $-1$,
respectively, and the other fields have baryon number zero. 

The generic Lagrangian terms for these hidden fields allowed by the
gauge symmetries are given by\footnote{We may also write a kinetic
mixing term between the U(1)$_Y$ and the hidden U(1) gauge fields, but
we suppress it in the present discussion. In fact, this kinetic mixing
is severely restricted by a recent global fit of solar precision data
so that $\chi_{\rm kin} m_{A^\prime} \lesssim 2\times 10^{-12}$~eV where
$\chi_{\rm kin}$ is the kinetic-mixing parameter and $m_{A^\prime}$ is
the mass of dark photon \cite{Vinyoles:2015aba, Schwarz:2015lqa}. The
absence of the kinetic  mixing can easily be explained if the SM
U(1)$_Y$ gauge group is embedded into a simple gauge group such as SU(5)
at high energies. In this case, the kinetic mixing is forbidden by the
gauge symmetry, and if there is no particle which is charged under both
the U(1)$_Y$ and the hidden U(1) gauge symmetries, then the kinetic
mixing is not generated even after the unified gauge symmetry is
spontaneously broken. } 
\begin{align}
 {\cal L}_{\rm hid} &= \sum_{i=1,2} \Psi^\dagger_i \overline{\sigma}^\mu i
 {\cal D}_\mu \Psi_i + \sum_{i=1,2} \overline{\Psi}^\dagger_i 
\overline{\sigma}^\mu i {\cal D}_\mu \overline{\Psi}_i 
+\overline{\chi}\left(i\slashed{\mathcal{D}}-m_\chi\right)\chi
+\frac{1}{2}\partial _\mu S \partial ^\mu S \nonumber \\
& -\left\{ y \overline{\chi}_R \chi_L S + {\rm h.c.}  \right\}
- V_{\rm sca} ~,
\label{eq:lag} 
\end{align}
with 
\begin{align}
 V_{\rm sca} = \frac{1}{2}m_S^2 S^2 
+\left(\mu_{S \Phi} S + \lambda_{S\Phi} S^2 \right)\Phi^\dagger \Phi 
+ \xi_S S + \frac{\kappa_S}{3!} S^3 + \frac{\lambda_S}{4!} S^4 
~,
\label{eq:pot}
\end{align}
where $\Phi$ is the SM Higgs doublet and $\mathcal{D}$s are the
covariant derivatives. We can always take the mass term of the DM
$\chi$ to be real; then, $y$ is in general complex, but we also take it
to be real for simplicity. Notice that the mass terms of the hidden
quark fields, as well as their couplings to the real scalar field $S$,
are forbidden by the gauge symmetries for $N > 2$. In these cases, the
conservation of the hidden baryon number is also a consequence of the
gauge symmetries. On the other hand, for $N=2$, vector-like mass terms
such as $\Psi_1 \Psi_2$ are allowed by the gauge symmetries. In this
particular case, we use the U(1)$_B$ to forbid these mass terms.
In any cases, this setup assures the hidden quark fields to be
massless.

\subsection{Confinement}

In the early Universe, the chiral fermions $\Psi_i$ and $\overline{\Psi}_i$
as well as the U(1) and SU($N$) gauge bosons act as massless elementary
fields and hence as DR. At later epochs, the non-Abelian SU($N$) gauge
interaction could confine if its coupling becomes large enough at
low energies and the temperature of the Universe fell below the
confinement scale. Then, due to the confinement, the chiral fermions and
the SU($N$) gauge bosons can not be regarded as fundamental fields any
more; instead, composite states such as hidden hadrons appear as physical
states. In addition, once the condensate of the chiral fermions forms,
the chiral flavor symmetry of the Lagrangian \eqref{eq:lag} is
spontaneously broken, and the Nambu--Goldstone bosons associated
with these broken symmetries show up. If the confinement scale is low
enough, these particles still behave as DR around the CMB epoch. For later
convenience, we briefly review the strong dynamics in our model and
refer Ref.~\cite{Harigaya:2016rwr} for detailed discussions, where the
hidden charged pion was regarded as a DM candidate.\footnote{The
possibility of the hidden charged pion being DR was pointed out in
Ref.~\cite{Co:2016akw}. For previous studies in which (elementary)
Nambu--Goldstone bosons are considered as DR, see
Refs.~\cite{Nakayama:2010vs, Lindner:2011it, Weinberg:2013kea,
Kawasaki:2015ofa}, and DR in other scenarios~\cite{Fischler:2010xz, Hooper:2011aj, Boehm:2012gr, Kelso:2013nwa, Geng:2013oda, Ko:2014bka, Ko:2016fcd, Ko:2016uft} for example.}

To that end, let us start with looking into the running of
the hidden SU($N$) gauge coupling ${g}_N$. The running of the
SU($N$) gauge coupling at one-loop level is given by
\begin{equation}
 \mu \frac{d {g}_N}{d \mu} = - \frac{{g}_N^3}{(4\pi)^2} 
\left(\frac{11}{3} N - \frac{4}{3}\right) ~,
\label{eq:rge}
\end{equation}
where $\mu$ is the renormalization scale. Therefore, for any $N \geq 2$,
the SU($N$) gauge theory is asymptotically free and its gauge coupling
becomes very strong in the infrared region. The confinement scale
$\Lambda$ for this gauge theory is estimated as
\begin{equation}
\Lambda \simeq \mu _0 \exp \left[-\frac{8\pi^2}{{g}^2_N(\mu_0)} \times
		       \frac{3}{11N - 4} \right]~,
\label{eq:confine}
\end{equation}
where ${g}_N(\mu_0)$ is an input value of ${g}_N$ at a
scale $\mu_0$ where perturbativity still holds. As we discuss later in
Sec.~\ref{sec:drpheno}, the observation of the CMB anisotropy restricts
the confinement scale to be $\Lambda \lesssim 1$~eV in this model. Such
a low confinement scale can easily be obtained with an ${\cal O}(1)$
input value of ${g}_N (\mu_0)$; for instance, for $N=2$, we have
$\Lambda \lesssim 1$~eV if $g_N (\mu_0) \lesssim 0.66$ (0.46) for
$\mu_0 = 10$~TeV (the Planck scale $M_P = 2.4\times 10^{18}$~GeV).

Since the hidden quarks in our model are massless, there is a global flavor
symmetry. In the absence of the U(1) gauge symmetry, the flavor symmetry
is maximal: $\text{SU}(2)_L \otimes \text{SU}(2)_R \otimes {\rm U}(1)_V$. The
axial U(1) symmetry is explicitly broken by anomalies. Once the U(1)
gauge symmetry is turned on, a part of this flavor symmetry is explicitly
broken. As we see below, however, we take the U(1) gauge coupling
${e}_D$ to be as small as $10^{-(3-4)}$, and thus the flavor-symmetry
breaking effects from the U(1) gauge interactions can be treated as a
small perturbation.

Now suppose that below the confinement scale $\Lambda$ the hidden quarks
condense such that the $\text{SU}(2)_L \otimes \text{SU}(2)_R$ flavor
symmetry is spontaneously broken into the ``isospin'' subgroup
$\text{SU}(2)_V$ just like the ordinary QCD; namely, the scalar bilinear
of the hidden quarks develops a vacuum expectation value of
\begin{align}
 \langle \overline{\Psi} \Psi \rangle \equiv 
 \langle \Psi_1 \overline{\Psi}_1 + \Psi_1^\dagger
 \overline{\Psi}_1^\dagger \rangle 
=
 \langle \Psi_2 \overline{\Psi}_2 + \Psi_2^\dagger
 \overline{\Psi}_2^\dagger \rangle  
\neq 0 ~,
\label{eq:vev}
\end{align}
where we expect $\langle \overline{\Psi} \Psi \rangle \sim
\Lambda^3$. Then, we obtain three Nambu--Goldstone bosons associated
with the broken axial-vector subgroup SU(2)$_A$. We refer to these
fields as the hidden pions (or dark pions) and denote them
by $\pi^{\prime a}$ ($a = 1,2,3$) together with $\pi^{\prime 0} \equiv
\pi^{\prime 3}$ and $\pi^{\prime \pm} \equiv (\pi^{\prime 1}\pm i
\pi^{\prime 2})/\sqrt{2}$. Notice that the hidden U(1) charge commutes
with the generator of SU(2)$_A$ corresponding to the neutral hidden pion
$\pi^{\prime 0}$---for this reason, $\pi^{\prime 0}$ is exactly
massless. On the other hand, the charged hidden pions $\pi^{\prime \pm}$
in general acquire a mass through the radiative corrections by the U(1)
gauge boson, which is estimated as 
\begin{equation}
 m_{\pi^{\prime \pm}}^2 \sim \frac{e_D^2}{16\pi^2} Q_\Psi (Q_\Psi -2)
  m_{\rho^\prime}^2 ~,
\end{equation}
where $m_{\rho^\prime}$ is the mass of the ``hidden $\rho$'', which is
expected to be around the cut-off scale of the effective theory of the
hidden pions, namely, $m_{\rho^\prime} \sim 4\pi f_{\pi^\prime}$ where
$f_{\pi^\prime} \sim \Lambda$ is the ``hidden-pion decay constant''. 
See Refs.~\cite{Das:1967it, Ecker:1988te, Harigaya:2016rwr} for more
careful estimations.

The condensate $\langle \overline{\Psi} \Psi \rangle$ in
Eq.~\eqref{eq:vev} has non-zero U(1) charge, and thus breaks the hidden
U(1) gauge symmetry as well. The hidden U(1) photon $A^\prime$ (or dark
photon) eats the neutral hidden pion $\pi^{\prime 0}$ to become massive,
and its mass $m_{A^\prime}$ is approximately given by
\begin{equation}
 m_{A^\prime} \sim 2 {e}_D f_{\pi^\prime} ~.
\end{equation}
If $e_D f_{\pi^\prime} \sim {e}_D \Lambda \lesssim 0.1$~eV, then this
hidden U(1) photon, together with $\pi^{\prime \pm}$, behaves as DR
around the CMB epoch.

Other hadronic states, such as ``hidden $\eta$'', ``hidden
$\rho$'', ``hidden baryons'', and so on, have masses of the order of the
cut-off scale of the low-energy effective theory for hidden
pions. The heavy hidden mesons can rapidly decay into hidden pions
or dark photons, and thus play no role in the following analysis. 
On the other hand, ``hidden nucleons'' are stable due to the hidden
baryon number.\footnote{The hidden baryon number is anomalous under the
hidden U(1) gauge interaction, but this symmetry-breaking effect on the
stability of hidden nucleons is negligibly small.} 
Hence, hidden nucleons can
potentially be DM in the Universe. Nevertheless, their thermal relic
abundance is extremely small since they can efficiently annihilate into
hidden pions through an ${\cal O}(1)$ ``pion-nucleon coupling''. As a
consequence, we can safely neglect their contribution to the
cosmological evolution in the following discussion.

\subsection{Dark matter sector}

DM in our model couples to the SM sector only through the real
singlet scalar $S$, which mixes with the SM Higgs field via the
trilinear coupling $\mu_S$. This setup is the same as the so-called
fermionic Higgs portal DM model~\cite{Kim:2008pp, Okada:2010wd,
Baek:2011aa, LopezHonorez:2012kv, Baek:2012uj}. Our model however has a
new intriguing feature which is absent in the simple fermionic Higgs
portal DM model. At Lagrangian level, the DM is stable because of the
U(1) gauge symmetry. This stability is not spoiled even after the U(1)
symmetry is spontaneously broken by the hidden-fermion condensate since
there remains a $\mathbb{Z}_2$ symmetry, which is a subgroup of the
hidden U(1) gauge symmetry.

To see this feature, let us consider the following transformation: 
\begin{align}
 S&\to S, \qquad \chi\to e^{i\pi} \chi, \qquad
\Psi_1 \to e^{i Q_\Psi \pi} \,\Psi_1, \qquad
\Psi_2 \to e^{-i Q_\Psi \pi}\, \Psi_2,
\nonumber \\[2pt]
\overline{\Psi}_1 &\to  e^{-i (Q_\Psi-2) \pi}\,\overline{\Psi}_1,\qquad
\overline{\Psi}_2 \to e^{i (Q_\Psi-2) \pi}\,\overline{\Psi}_2~.
\label{eq:transf}
\end{align}
The Lagrangian \eqref{eq:lag} is invariant under this transformation, 
which follows from
the hidden U(1) gauge symmetry. In addition, the hidden quark condensate
\eqref{eq:vev} is also invariant under this transformation. Therefore,
the transformation \eqref{eq:transf}, which is a subgroup of the U(1)
gauge symmetry, remains a good symmetry even after the hidden U(1) gauge
symmetry is spontaneously broken. Meanwhile, the transformation
\eqref{eq:transf} is nothing but a $\mathbb{Z}_2$ symmetry under which
the DM $\chi$ is odd while the other particles are even. As a
consequence, we find that the condensate \eqref{eq:vev} breaks the U(1)
symmetry down to a $\mathbb{Z}_2$ symmetry \cite{Krauss:1988zc} which
stabilizes the DM particle.\footnote{For DM models which exploit such a
remnant discrete symmetry, see Refs.~\cite{Kadastik:2009dj,Kadastik:2009cu, 
Frigerio:2009wf, Hambye:2010zb, Mambrini:2013iaa,Baek:2014kna,Ko:2014nha,
Mambrini:2015vna, Heeck:2015qra, Nagata:2015dma, Nagata:2016knk}.}
Thanks to this $\mathbb{Z}_2$ symmetry, the stability of DM is insured
even if there exist higher-dimensional operators induced by ultraviolet
effects.

\subsection{Scalar sector}

The connection between the dark sector and the SM particles is provided 
through the Higgs-portal terms, $\left(\mu_S S +
\lambda_\phi S^2 \right)\Phi^\dagger \Phi$. If $\mu_S\neq
0$,\footnote{We here assume that the singlet field $S$ does not develop
a vacuum expectation value, just for simplicity. Relaxing this
assumption does not change our discussion so much. } there is a mixing
between $S$ and the Higgs boson $\phi$ after the electroweak symmetry is
broken, where $\Phi = (v+\phi)/\sqrt{2}$ with $v\simeq 246$~GeV. As a
result, $\chi$ can couple to the SM Higgs boson and thus to all of the
SM particles via the mixing. The mass matrix for $\phi$ and $S$ in the
$(\phi, S)$ basis is given by 
\begin{equation}
\mathcal{M}^{2}=
\begin{pmatrix}
2\lambda_{\Phi}v^{2} & \mu_{S}v\\
\mu_{S}v &  m^2_S+\lambda_S v^2
\end{pmatrix}
~,
\end{equation}
where $\lambda_\Phi$ is the quartic coupling in the SM Higgs potential:
$\lambda_\Phi \left(\Phi^\dagger \Phi\right)^2$. Diagonalization of the
above mass matrix results in two mass eigenstates $h$ and $s$: 
\begin{equation}
\left(\begin{array}{c}
h\\
s
\end{array}\right)
=\left(\begin{array}{cc}
\cos{\alpha} & {}-\sin{\alpha}\\
\sin{\alpha} & \cos{\alpha}
\end{array}\right)\left(\begin{array}{c}
\phi\\
S
\end{array}\right)~,
\end{equation}
where the mixing angle $\alpha$ is given by
\begin{equation}
 \tan{2\alpha}=\frac{2\mathcal{M}_{12}^{2}}{\mathcal{M}_{22}^{2}-
\mathcal{M}_{11}^{2}}
=\frac{\mu_{S}v}{m^2_S+\lambda_S v^2-2\lambda_{\Phi}v^{2}} ~.
\end{equation}
The mass eigenvalues of $h$ and $s$ are
\begin{equation}
m_{h,s}^{2}=\lambda_{\Phi}v^{2}+\frac{m^2_S+\lambda_S v^2}{2}
 \pm\sqrt{\left(\lambda_{\Phi}v^{2}-\frac{m^2_S+\lambda_S
	   v^2}{2}\right)^{2}+\left(\mu_S v\right)^{2}} ~.
\label{eq:H1H2mass}
\end{equation}
Using these masses, the mixing angle can also be given as
\begin{equation}
 \sin{2\alpha}=\frac{2\mu_S v}{m_{s}^{2}-m_{h}^{2}} ~.
\end{equation}
In the rest of our discussion, we shall identify $h$ as the Higgs boson
with $m_{h}\simeq 125~\GeV$ and treat $m_{s}$ and other parameters as
free variables.

If the mass of $s$ is less than a half of the Higgs boson mass, we would
have exotic decay channels of the Higgs boson such as $h\rightarrow
s+s\rightarrow 4f$. No observation of such signals then puts constraints
on the parameters $\mu_S$ and $\lambda_{S}$. The current bound
\cite{Aaboud:2016oyb, Liu:2016ahc} can easily be satisfied if
$\mu^2_S/v^2\lesssim 10^{-3}$ and  $\lambda_S\lesssim 10^{-3}$. This
bound is of course evaded if $m_s > m_h/2$. The results of the Higgs
boson measurements at the LHC also give a constraint on the mixing angle
$\alpha$, but it is still rather weak; $|\alpha| \lesssim 0.1$ is enough
to evade all of the existing bounds from the Higgs data \cite{Cheung:2015dta}.

\section{Thermal History}
\label{sec:pheno}

\subsection{Thermalization of the dark sector}
\label{sec:thermds}

\begin{figure}[t]
\centering
\includegraphics[width=0.95\textwidth]{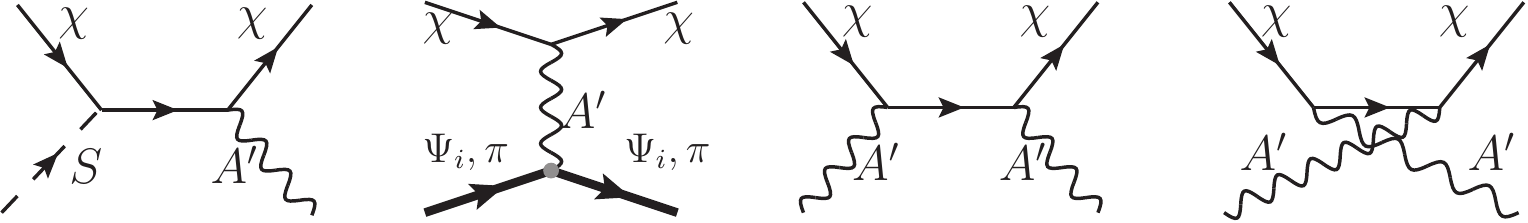}
\vspace*{3mm}~\\
\includegraphics[width=0.95\textwidth]{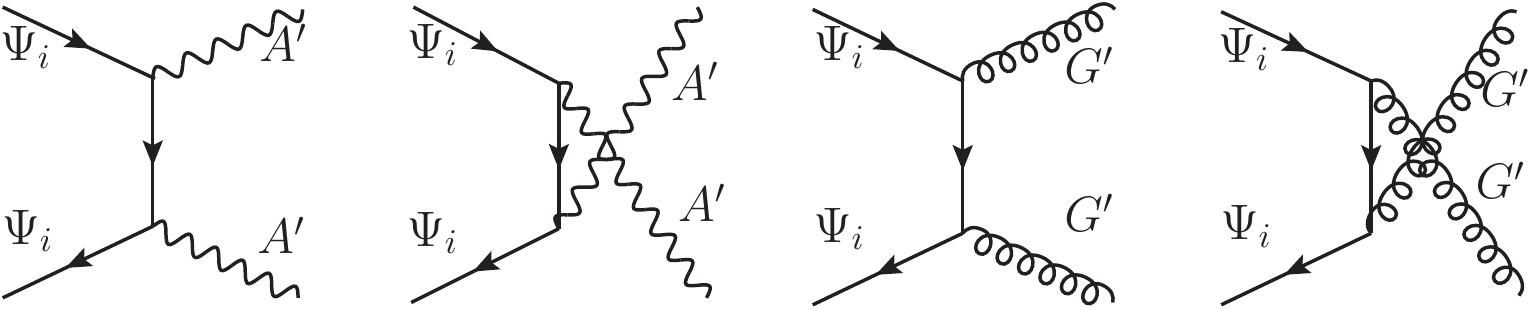}
\caption{Typical scattering processes in the dark sector. Upper panel:
 scattering between DM $\chi$ and other hidden-sector particles. Lower
 panel: thermalization processes among light particles in the dark
 radiation sector. } 
\label{fig:feynmann}
\end{figure}

In this section, we discuss the thermal history of the present
model. The dark sector in our model is assumed to be in thermal
equilibrium with the SM sector in the early Universe. This can be
realized through the Higgs portal couplings, and this requirement
imposes lower limits on these couplings. To have the dark sector in
thermal equilibrium with the SM sector, the scattering rate $\Gamma$
for relevant processes such as $\Phi \Phi^\dagger \leftrightarrow SS$
should be larger than Hubble parameter $H$,  
\begin{equation}
\Gamma\sim \left(\lambda^2_S+\frac{\mu^2_S}{v^2}\right)T \gtrsim H \sim
 \frac{T^2}{M_P}~, 
\end{equation} 
where $T$ is the thermal temperature and $M_P\simeq 2.4\times
10^{18}$~GeV is the reduced Planck mass. From this expression, one can
see that this condition is satisfied at a late time. For this to happen
at a temperature $T > m_\chi \gtrsim 1$~TeV, therefore, we need 
\begin{equation}
 \left(\lambda^2_S+\frac{\mu^2_S}{v^2}\right) 
\gtrsim \left(\frac{m_\chi}{M_P}\right) = 
2 \times 10^{-8} \times
\left(\frac{m_\chi}{1~{\rm TeV}}\right)
~.
\end{equation}
As long as this condition is satisfied, the singlet scalar $S$ is in
thermal equilibrium with the SM sector at a high temperature. Once this
occurs, DM $\chi$ is also thermalized via the Yukawa coupling $y$,
and then the scattering processes shown in Fig.~\ref{fig:feynmann} take
the whole dark sector in thermal equilibrium.

\subsection{Relic density of dark matter}

\begin{figure}[t]
\centering
\includegraphics[width=0.8\textwidth]{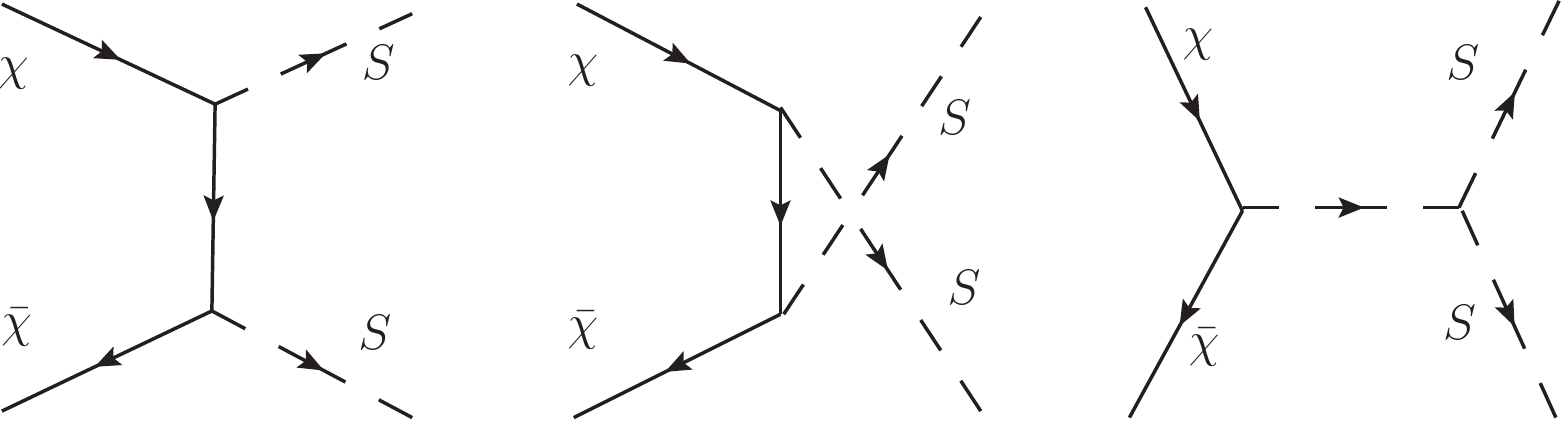}
\caption{Annihilation processes for $\chi \bar{\chi}\rightarrow S S$.  }
\label{fig:annih}
\end{figure}

Next, we evaluate the thermal relic abundance of DM $\chi$. Here, we
focus on the case where $m_s < m_\chi$, though it is not a necessary
condition. In this case, the relic density of DM $\chi$ is mainly
determined by the annihilation process, $\overline{\chi}+\chi\rightarrow
S + S$, shown in Fig.~\ref{fig:annih}. For simplicity, we neglect the
contribution from the last diagram due to the ignorance of triple-scalar
coupling $\kappa_S$. In the case where $m_s$ is much less than $m_\chi$,
we can estimate the annihilation cross section of $\chi$ as
\begin{equation}
\langle \sigma v_{\rm rel}\rangle \sim \frac{y^4T}{16\pi^2m^3_\chi} ~,
\label{eq:relic}
\end{equation}
where $v_{\rm rel}$ is the relative velocity between the annihilating DM
particles and $T\simeq m_\chi/20$ at the freeze-out time. To get
the correct relic density $\Omega_{\rm DM} h^2 \simeq 0.12$
\cite{Planck:2015xua}, we need $\langle \sigma v_{\rm rel}\rangle \simeq 3\times
10^{-26}$~cm$^{3}/$s, which basically fixes the relation between the
Yukawa coupling $y$ and the DM mass $m_\chi$. Since the annihilation
processes are $p$-wave suppressed, the annihilation cross section in the
current Universe is extremely small, and thus constraints from DM
indirect searches can easily be avoided.
We also note that the Sommerfeld enhancement \cite{Hisano:2003ec,
Hisano:2004ds} for the DM annihilation due to the new U(1) gauge
interaction can be neglected thanks to the smallness of the gauge
coupling ${e}_D$. This can be seen from the enhancement
factor~\cite{ArkaniHamed:2008qn, Feng:2009mn}  
\begin{equation}
 F=\frac{\pi {\alpha}_D/v_{\rm rel}}{1-e^{-\pi {\alpha}_D/v_{\rm rel}}}~,
\end{equation}
with ${\alpha}_D\equiv {e}_D^2/4\pi$. As we mentioned above, we take
${e}_D \sim 10^{-(3-4)}$, {\it i.e.}, ${\alpha}_D \sim
10^{-(7-9)}$. Since this is much smaller than $v_{\rm rel}\simeq 10^{-3}$ (a
typical size of DM velocities in galaxies), we have $F\simeq 1$, which would
not change the annihilation cross section drastically.

Although indirect DM searches are less promising, DM direct detection
experiments may probe the DM candidate in our model. The DM-nucleon
scattering process is induced by the exchange of the scalar bosons $h$
and $s$, and its spin-independent scattering cross section is given by
\cite{Shifman:1978zn} 
\begin{equation}
 \sigma_{\rm SI}^{(N)} = \frac{f_N^2}{\pi} \frac{m_N^2 m_\chi^2}{(m_N +
  m_\chi)^2} ~,
\end{equation}
with 
\begin{equation}
 \frac{f_N}{m_N} = \frac{y}{2v} \sin 2\alpha \left(\frac{1}{m_h^2}
- \frac{1}{m_s^2}\right)
\left[\sum_{q=u,d,s} f_{Tq}^{(N)} + \frac{2}{9}f_{TG}^{(N)}\right] ~,
\end{equation}
where $m_N$ is the nucleon mass, $f_{Tq}^{(N)} \equiv \langle N | m_q
\overline{q}q |N\rangle /m_N$ are the mass fractions, and $f_{TG}^{(N)}
\equiv 1-\sum_{q} f_{Tq}^{(N)}$. These mass fractions are computed using
lattice QCD simulations \cite{Abdel-Rehim:2016won}: $f_{Tu}^{(p)} = 0.0149$,
$f_{Td}^{(p)} =0.0234$, and $f_{Ts}^{(p)} = 0.0440$ for
proton. According to the rough estimate given in Eq.~\eqref{eq:relic},
the correct DM density is obtained for, {\it e.g.}, $y \simeq 0.36$ and
$m_\chi \simeq 1$~TeV. In this case, we obtain $\sigma_{\rm SI}^{(p)}
\simeq 5 \times 10^{-46}$~cm$^2$ for $m_s = 300$~GeV and $\tan \alpha =
0.1$. This size of $\sigma_{\rm SI}^{(p)}$ evades the current 
experimental bound provided by the XENON1T experiment
\cite{Aprile:2017iyp}, but is within the reach of future DM direct
detection experiments such as a two-year measurement at XENON1T
\cite{Aprile:2015uzo}, and therefore we may probe this 
scenario in the near future. More dedicated studies on the fermionic
Higgs-portal DM scenario will give further prospects for the testability
of this scenario in future experiments (see, for instance,
Ref.~\cite{Esch:2013rta} for a recent study on the detectability of the
fermionic Higgs-portal DM).

\subsection{Decoupling of the dark sector}

As we have seen in Sec.~\ref{sec:thermds}, the scattering process
$\chi+S \leftrightarrow \chi +A^\prime$ can keep the dark photon
$A^\prime$ in thermal equilibrium. When the temperature becomes lower
than the DM mass, however, the rate of this process gets suppressed as the DM
number density exponentially decreases. Eventually, $A^\prime$ decouples
at a temperature $T_{\rm dec}$. Since the hidden U(1) gauge
coupling is taken to be much smaller than the Yukawa coupling $y$,
$A^\prime$ decoupled earlier than $\chi$, {\it i.e.}, $T_{\rm dec} >
m_\chi/20$. Indeed, we can estimate the decoupling temperature 
$T_{\rm dec}$ by comparing the scattering rate for $\chi+S
\leftrightarrow \chi +A^\prime$ with the Hubble expansion rate:
\begin{equation}
 n_\chi (T_{\rm dec}) \cdot \frac{y^2e_D^2}{16\pi^2 m_\chi^2} 
\sim \frac{T_{\rm dec}^2}{M_P}~,
\end{equation}
where $n_\chi (T)$ denotes the number density of the DM particle $\chi$,
which is given by 
\begin{equation}
 n_\chi(T) = 4\biggl(\frac{m_\chi 
T}{2\pi}\biggr)^{3/2}\textrm{exp}\biggl[-\frac{m_\chi}{T}\biggr]~.
\end{equation}
This leads to 
\begin{align}
 \frac{m_\chi}{T_{\rm dec}} &\sim \ln \biggl[
\frac{M_P}{\sqrt{m_\chi T_{\rm dec}}}
\frac{y^2e_D^2}{4\pi (2\pi)^{\frac{3}{2}}}
\biggr] \nonumber \\[3pt]
&\sim 10 +2\ln \biggl(\frac{ye_D}{10^{-4}}\biggr)
 -\ln\biggl(\frac{m_\chi}{1~{\rm TeV}}\biggr) 
+\frac{1}{2}\ln \biggl(\frac{m_\chi}{T_{\rm dec}}\biggr) ~.
\end{align}
This estimation shows that up to the small logarithmic dependence on
parameters the decoupling temperature $T_{\rm dec}$ in the present
scenario is given by $T_{\rm dec}/m_\chi \sim 1/10$. Therefore, if the
DM mass is ${\cal O}(1)$~TeV, then $T_{\rm dec} = {\cal O}(100)$~GeV.
Other scattering processes between DM and the dark photon
or the hidden quarks (corresponding to the diagrams in the upper row in
Fig.~\ref{fig:feynmann}) are further suppressed by the small U(1) gauge
coupling ${e}_D$, and thus decoupled earlier. Consequently, the
DR sector decouples from the SM sector at the temperature $T_{\rm dec}$, which
is before the chemical decoupling of DM. On the other hand, the
processes $A^\prime +A^\prime \leftrightarrow \Psi_i +
{\Psi}^\dagger_i$, $ \Psi_i + {\Psi}_i^{\dagger} \leftrightarrow
G^\prime + G^\prime$, and $ \overline{\Psi}_i +
\overline{\Psi}_i^{\dagger} \leftrightarrow G^\prime + G^\prime$ with
$G^\prime$ being the SU($N$) gauge boson, which are described by the
diagrams in the lower row in Fig.~\ref{fig:feynmann}, always keep those
species in equilibrium with each other, especially at low energies. This
is because the scattering rate for $A^\prime +A^\prime \leftrightarrow
\Psi_i + {\Psi}^\dagger_i$ goes as $\sim e^4_D T$, in comparison with
Hubble parameter $T^2/M_P$. This DR sector is composed of free hidden
quarks, dark gluons $G^\prime$, and dark photons $A^\prime$ before the
confinement of the SU($N$) gauge interaction. Below the confinement
scale $\Lambda$, only the dark pions and dark photons are left in the
cosmic background.

\begin{figure}[t]
\centering
	\includegraphics[width=0.6\textwidth]{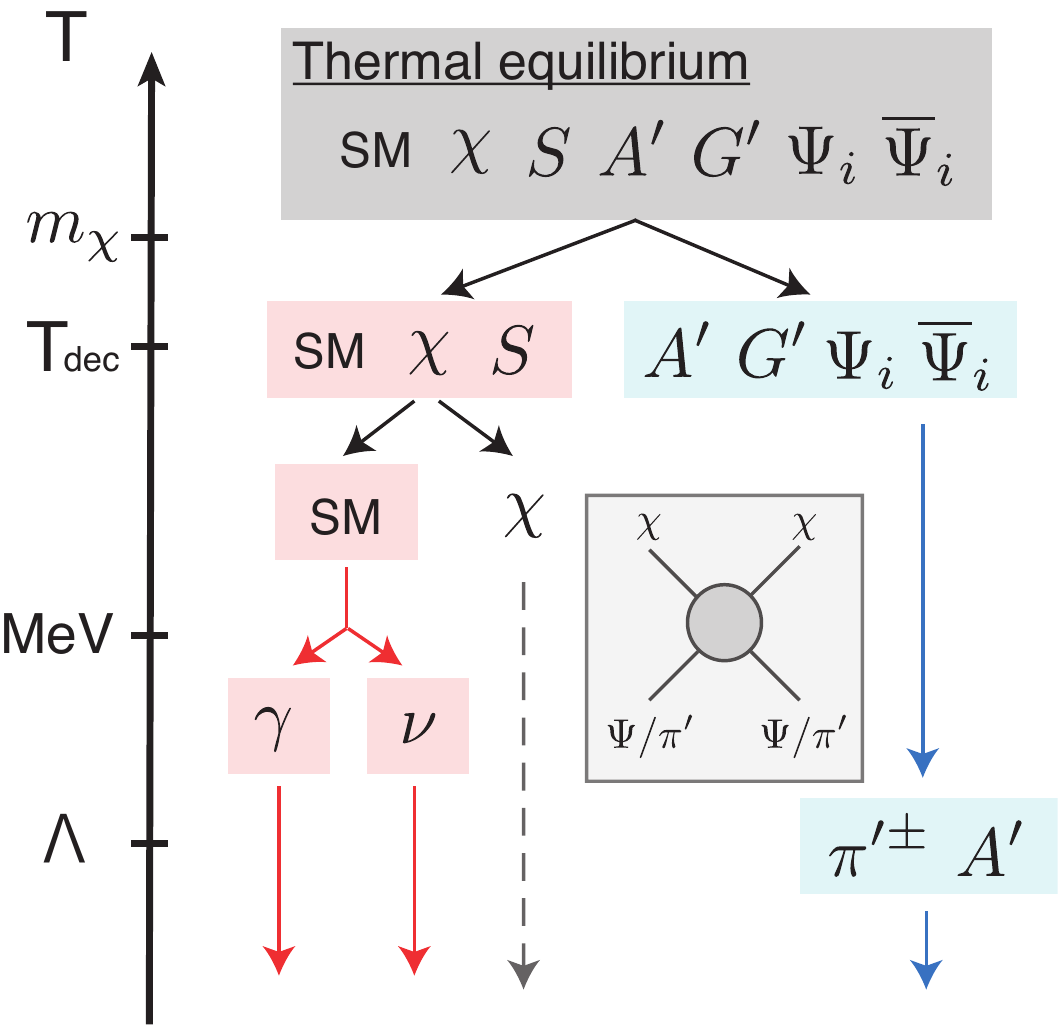}
	\caption{Thermal history of our model.}
	\label{fig:thermal}
\end{figure}

After all, the thermal history in this model goes as follows. In the
early Universe, the Higgs-portal couplings keep the whole dark sector in
equilibrium with the SM sector. When temperature becomes as low as
$T_{\rm dec} < m_\chi$, $\Psi_i$, $\overline{\Psi}_i$, $A^\prime$, and
$G^\prime$ decouple from the SM sector. Then, the chemical decoupling of
$\chi$ occurs at a temperature $T \simeq m_\chi/20 < T_{\rm dec}$ and its
abundance freezes out. $S$ is not stable and decays into the SM
sector. After that, the SM and DR sectors are independently thermalized
in each sector. The DM $\chi$ scatters with light particles in the DR
sector via the exchange of a dark photon---with the hidden
charged quarks for $T > \Lambda$ and with the hidden charged pions
$\pi^{\prime \pm}$ for $T < \Lambda$. This continues until the time of
the matter-radiation equality. The overall picture for the
thermal history is illustrated in Fig.~\ref{fig:thermal}.

\section{Dark Radiation and Diffusion Damping}
\label{sec:drpheno}

Now we discuss the phenomenological consequences of DR in our model. In
Sec.~\ref{sec:drab}, we evaluate the contribution of DR to the effective
number of neutrinos, $N_\eff$. We then discuss the effects of the DM-DR
interactions on the matter power spectrum of DM in
Sec.~\ref{sec:diffdamp}. 

\subsection{Dark radiation}
\label{sec:drab}

The hidden quarks, gluons, and photons (or the hidden pions and photons
below the confinement scale) behave as DR, and thus contribute to the
effective number of neutrinos. However, this contribution is fairly
suppressed since the DR sector decouples much before the decoupling of
neutrinos. Taking this suppression into account, we compute the
shift in $N_\eff$ caused by the DR as 
\begin{align}
\delta  N_{\textrm{eff}}  = \left(\frac{8}{7} N_b
+ N_f\right)\frac{T^4_D}{T^4_\nu}=\left(\frac{8}{7} N_b
+ N_f\right)\left[ \frac{g_{\ast s}\left(T_{\nu,
 \textrm{dec}}\right)}{g_{\ast s}\left(T_{\textrm{dec}}\right)} 
\frac{g_{\ast s}^{D} \left(T_{\textrm{dec}}\right)}{g_{\ast
 s}^{D}\left(T_{D}\right)} \right]^{\frac{4}{3}},
\end{align}
where $N_b$ and $N_f$ are the bosonic and fermionic degrees of freedoms,
normalized to massless gauge boson and Weyl fermion, respectively.
$T_\nu$ and $T_D$ are the temperature of neutrinos and the DR sector,
respectively, $T_{\nu, {\rm dec}}$ is the neutrino decoupling
temperature, $g_{\ast s} (T)$ denotes the effective number of degrees of
freedom for entropy density in the SM sector at temperature $T$, and
$g_{\ast s}^{D}$ denotes the effective number of degrees of freedom that
are in kinetic equilibrium with dark photon. In the last equality, we
have used the conservation of entropy density.

A feature of our model is that $\delta  N_{\textrm{eff}}$ could change
over the time due to the factor $g_{\ast s}^{D}
\left(T_{\textrm{dec}}\right)/g_{\ast s}^{D}\left(T_{D}\right)$. This is
because the physical degrees of freedom in the hidden sector change when
the temperature falls down below the confinement scale. Above the
confinement scale, we have  
\begin{align}
\delta  N_{\textrm{eff}}  = \left[\frac{8}{7}\times \left\{(N^2-1) +1
\right\}  
+ 4 \times N\right]\times\frac{T^4_D}{T^4_\nu}
 \simeq \frac{4N(2N+7)}{7}
\left[ \frac{g_{\ast s}\left(T_{\nu, {\rm dec}}\right)}{g_{\ast
		s}\left(T_{\textrm{dec}}\right)} \right]^{\frac{4}{3}}.
\label{eq:geff1}
\end{align}
Therefore, if $T_{\textrm{dec}}\gg m_t\simeq 173~\GeV$, we
obtain $\delta  N_{\textrm{eff}}$ at the BBN epoch as\footnote{Here,
we have assumed that $S$ behaves as a non-relativistic particle at $T =
T_{\rm dec}$. } 
\begin{equation}
\delta  N_{\textrm{eff}} = \frac{4N(2N+7)}{7}
\left[ \frac{43/4}{427/4}\right]^{\frac{4}{3}}\simeq 
0.047 \times \frac{4N(2N+7)}{7} ~.
\end{equation}
This leads to $\delta N_{\rm eff} = 0.59$ and $1.04$ for $N=2$ and 3,
respectively. Below the confinement scale, on the other hand, we have
only hidden pions and photons, and thus
\begin{align}
\delta  N_{\textrm{eff}}  = \left[\frac{8}{7}\times
 \left(\frac{3}{2}+1\right)\right]\times 
\left[ \frac{g_{\ast s}\left(T_{\nu, {\rm dec}}\right)}{g_{\ast
		s}\left(T_{\textrm{dec}}\right)} \dfrac{N(2N+7)}{5}\right]^{\frac{4}{3}},
\label{eq:geff2}
\end{align}
where we have used $g_{\ast s}^{D}
\left(T_{\textrm{dec}}\right)=N(2N+7)$ ($5$) above (below) the
confinement scale, which is obtained by multiplying the prefactor in
Eq.~\eqref{eq:geff1} (Eq.~\eqref{eq:geff2}) by a factor of $7/4$. Again,
for $T_{\textrm{dec}}\gg m_t\simeq 173~\GeV$, we have 
\begin{equation}
\delta  N_{\textrm{eff}}\simeq 0.134\times \left[
\frac{N(2N+7)}{5}\right]^{\frac{4}{3}},  
\end{equation}
which would give $\delta  N_{\textrm{eff}}=0.97$ for $N=2$ and $\delta
N_{\textrm{eff}}=2.1$ for $N=3$. 
 
As a result, we find that the $N=2$ case provides a value of $\delta
N_{\rm eff}$ which lies in the 
favored range $ 0.4 \lesssim \delta N_{\rm eff} \lesssim 1$ to relax the
tension in observed values of $H_0$ \cite{Riess:2016jrr}, while the
$N\geq 3$ case is disfavored. Future CMB experiments such as CMB-S4
\cite{Abazajian:2016yjj} may determine the value of $N_{\rm eff}$ within
an error of 0.02--0.03, and thus can test this scenario with great
accuracy since $\delta N_{\rm eff} \geq 0.59$ in this model.  

We note that \planck\cite{Planck:2015xua} gives an upper bound on
$\delta N_{\rm eff}\lesssim 0.7$, which relies on combinations of data
sets from different measurements and the assumed cosmological
models. Possible systematic uncertainties and extended cosmological
parameters could give more relaxed limits. Nevertheless, the above
bound, if robust, would constrain the confinement scale, $\Lambda
\lesssim 1$~eV so that the hidden chiral fermions and gluons, rather
than hidden pions, comprise physical degrees of freedom around the CMB
epoch. Such a confinement scale would ensure $\delta N_{\rm eff}\simeq
0.59$ during the CMB time, which evades the Planck limit. After the CMB
time, even if $\delta N_{\rm eff}$ increases to $0.97$, DR does not
affect the CMB anisotropy significantly since its contribution to the energy
density is by far smaller than that of the matter component.

\subsection{Diffusion damping}
\label{sec:diffdamp}

\begin{figure}[t]
 \centering
 \includegraphics[width=0.7\textwidth]{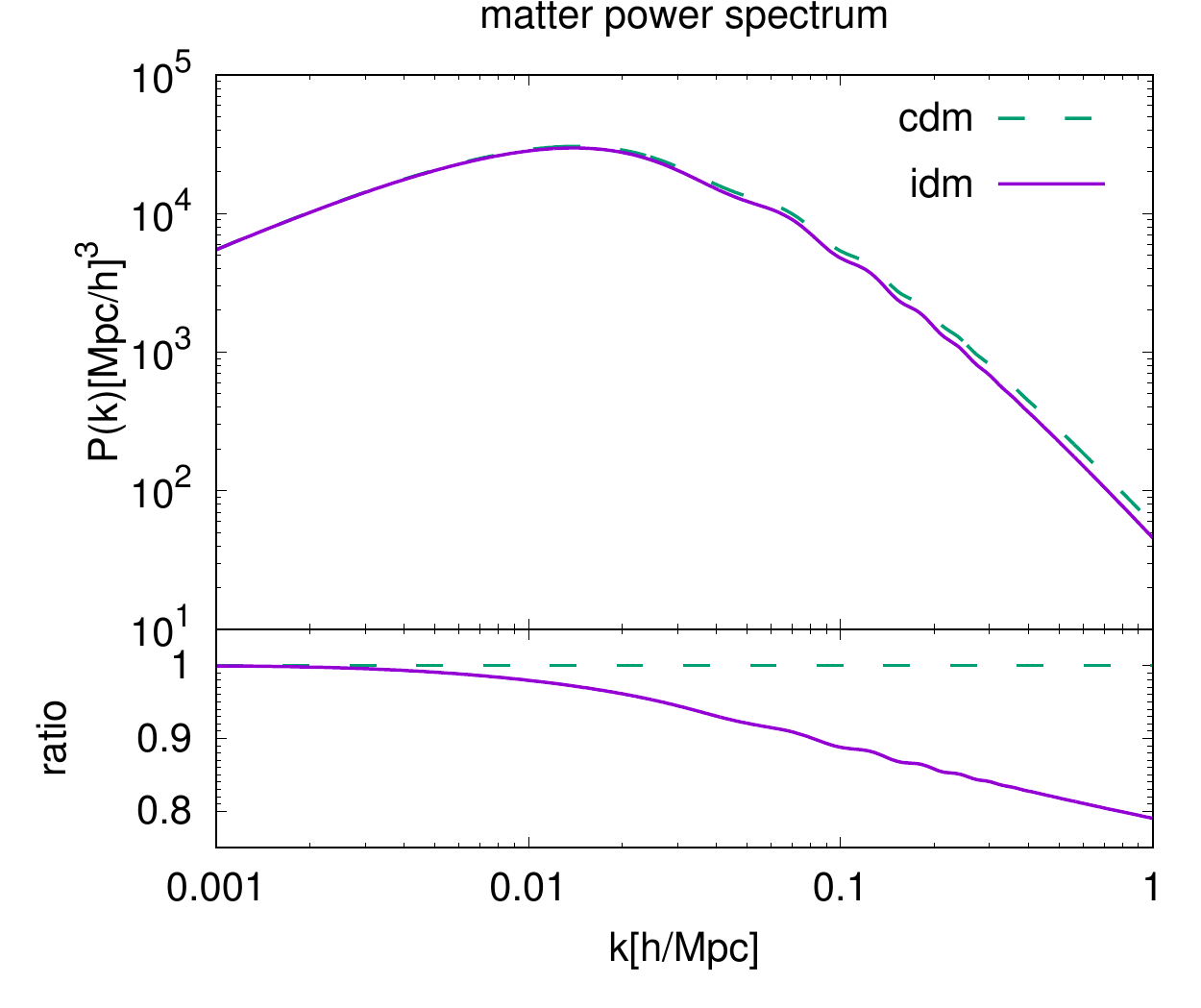}
 \caption{Matter power spectrum. The purple solid and green dashed lines
 show the matter spectrum with and without DM-DR interactions,
 respectively. }
 \label{fig:pk}
\end{figure}

The scattering between DM $\chi$ and hidden-charged particles can induce
diffusion damping in the matter power spectrum by modifying the
evolution of the DM density perturbation. To resolve the discrepancy in the observed values of $\sigma_8$, we need a size of the hidden U(1)
gauge coupling $e_D$ such that the interactions of $\chi$ and the U(1)
charged DR decouple around the radiation-matter equality
\cite{Lesgourgues:2015wza, Ko:2016uft}. This is estimated from the
condition \cite{Ko:2016uft}
\begin{equation}
n_{D}\cdot  \sigma_{\rm int}  \cdot \frac{T_D}{m_\chi} 
\simeq H ~,
\label{eq:deccond}
\end{equation}
where $n_D \sim T_D^3$ is the number density of DR and $\sigma_{\rm int}
\sim e_D^4/T_D^2$ is the typical size of the cross section of DM-DR
scatterings. It follows from this condition that
\begin{align}
e_D \sim \left(\frac{T_\gamma}{T_D}\right)^{\frac{1}{2}}
\left(\frac{m_\chi}{M_P}\right)^{\frac{1}{4}}
\simeq 1.4\times 10^{-4} \times 
\left(\frac{T_\gamma}{T_D}\right)^{\frac{1}{2}} \times
\biggl(
\frac{m_\chi}{1~{\rm TeV}}
\biggr)^{\frac{1}{4}}
~,
\label{eq:g1}
\end{align}
where $T_\gamma$ is the CMB temperature.

We need to include the scattering effects into the cosmological
evolution of perturbations, which we perform numerically. We modify the
Euler equations for DM $\chi$ and DR (collectively denoted as $dr$) to
\begin{align}
\dot{\theta}_\chi &= k^2\psi -\mathcal{H} \theta_\chi +
 S^{-1}\dot{\mu}\left(\theta_{dr}-\theta_{\chi}\right)~,\label{eq:veldiv1}\\[2pt] 
\dot{\theta}_{dr} &= k^2 \psi +k^2\left(\frac{1}{4}\delta_{dr}
 -\sigma_{dr}\right)-\dot{\mu}\left(\theta_{dr}-\theta_{\chi}\right)~ ,\label{eq:veldiv2}
\end{align}
where $k$ is the comoving wave number, $\psi$ is the gravitational
potential, the dot means derivative over conformal time $d\tau\equiv
dt/a$ ($a$ is the scale factor), $\theta_{dr}$ and $\theta_\chi$ are
velocity divergences of DR and DM $\chi$, respectively, 
$\delta_{dr}$ and $\sigma_{dr}$ are the density perturbation and the
anisotropic stress potential of DR, respectively, and $\mathcal{H}
\equiv \dot{a}/a$ is the conformal Hubble parameter. Finally, the
scattering rate and the density ratio are defined by $\dot{\mu}=a
n_{\chi}\langle\sigma_{\rm int}c\rangle$ and $S=3\rho_\chi/4\rho_{dr}$,
respectively. Note that $\sigma_{dr}=0$ in our interested parameter
regime because DR self-interacts strongly and behaves as a perfect
fluid. 

In Fig.~\ref{fig:pk}, we show the damping effects in our model, which
are obtained by using the Boltzmann code {\tt CLASS} \cite{class} with
implement of above perturbation equations. We take $\delta
N_{\eff}\simeq 0.59$ for the $N=2$ case, which gives $T_\gamma/T_D\simeq
2.15$ and $e_D$ is determined from the condition~\eqref{eq:deccond}. The
green dashed line shows the matter power spectrum for DM without DM-DR
interactions and the purple solid line is for our model. This figure
clearly shows the damping effects for wave-number $k\gtrsim
0.01$~$h/{\rm Mpc}$. With an $\mathcal{O}$(10--15)\% suppression at
$k\simeq 0.2$, we have $\sigma_8\simeq 0.74$, which is much closer to
the values obtained from weak-lensing measurements \cite{Heymans:2012gg}.

Based on what we have discussed above, we might also explore a variant
model where hidden SU($N$) is replaced by another U(1), in which case no
confinement or dark pion could arise. However, this variant model shares
similar features for resolving the cosmological tensions: 1) both
chiral fermions and gauge bosons are symmetry-assured massless and
contribute to DR. In this case, the anomaly cancellation condition
requires $\psi_1$ and $\Psi_2$, as well as $\overline{\Psi}_1$ and
$\overline{\Psi}_2$, to be vector-like, and their vector-like mass terms
are forbidden by the hidden baryon number. 2) DM scatters with DR, which
leads to a modified power spectrum. Since this model has less physical
degrees of freedom, the amount of DR can be reduced by a factor
$11/[N(2N+7)]$, in comparison to Eq.~\eqref{eq:geff1} for the SU($N$)
model, if decoupled at the same temperature. Another variant is to
introduce an extra set of chiral fermions. However, this case leads to a
larger value of $\delta N_{\rm eff}$ and thus is disfavored by the
Planck constraint. Finally, we may also consider the cases where the DM
$\chi$ is charged under the $\text{SU}(2)_L$ symmetry. The hypercharge
of this DM should be zero in order to suppress the vector coupling with $Z$
boson, which induces a too large DM-nucleon scattering cross section. In
this case, we do not need to introduce the singlet scalar $S$ to couple
the dark sector to the SM sector. A similar setup is considered in
Refs.~\cite{Buen-Abad:2015ova, Lesgourgues:2015wza}. The relic abundance
of such a particle agrees to the observed DM density if its mass is
${\cal O}(1-10)$~TeV depending on the SU(2)$_L$ charge
\cite{Hisano:2003ec, Cirelli:2005uq, Cirelli:2007xd}, which assures the
dark sector to decouple above the weak scale. Since the annihilation
cross sections of these DM candidates are rather large, they can
efficiently be probed in indirect detection experiments
\cite{Cohen:2013ama, Fan:2013faa, Hryczuk:2014hpa,
Bhattacherjee:2014dya, Baumgart:2014saa, Cirelli:2015bda,
Garcia-Cely:2015dda, Lefranc:2016fgn}. Their spin-independent scattering
cross sections with a nucleon are larger than the neutrino floor
background \cite{Hisano:2015rsa}, and thus they can also be tested in
future direct detection experiments.

\section{Conclusion and Discussions}
\label{sec:conc}

In this paper, we have illustrated a model where a fermionic DM particle
interacts with chiral/composite DR via the exchange of a hidden U(1)
gauge boson. The chiral DR, being massless assured by the symmetries in
this model, consists of the hidden SU($N$)-charged quarks. This sector
possesses a flavor symmetry that is spontaneously broken by the
hidden-quark condensate below the confinement scale
$\Lambda$ of the hidden SU($N$) gauge interaction. Then, dark pions as
the associated Nambu-Goldstone bosons become the DR below the
confinement scale. The hidden U(1) gauge symmetry is broken into a
$\mathbb{Z}_2$ symmetry by the hidden quark condensate, which assures
the stability of DM. The hidden U(1) gauge boson acquires a tiny mass
since the confinement scale is bounded as low as $\lesssim 1$~eV by the
limit on $N_{\rm eff}$. Thus, both the nearly massless charged
hidden pions and the dark photon behave as DR in our model. Thanks to
the early decoupling of the DR sector, the contribution of DR to the
effective number of neutrinos is $\delta N_{\rm eff}\simeq 0.59$ above
the confinement scale for the $N=2$ case, which can relax the tension in
the observed values of $H_0$ if $\Lambda \lesssim 1$~eV. The $N\geq 3$
cases are disfavored as $N_{\rm eff}$ is shifted 
too much. Moreover, the DM-DR interactions via the exchange of a dark
photon can induce diffusion damping in the matter power spectrum, which
accounts for the discrepancy in $\sigma_8$ obtained from the \planck and
other low red-shift measurements. The DM in our model couples to the SM
sector through the Higgs-portal coupling of a real singlet scalar, which
enables us to probe the DM in this model in future DM direct detection
experiments. We can also test this scenario in the next-generation CMB
experiments, such as CMB-S4 \cite{Abazajian:2016yjj}.

\section*{Acknowledgments}
The work of PK is supported in part by National Research Foundation of
Korea (NRF) Research Grant NRF-2015R1A2A1A05001869, and by the NRF grant
funded by the Korea government (MSIP) (No.~2009-0083526) through Korea
Neutrino Research Center at Seoul National University. This work is
supported by the Grant-in-Aid for Scientific Research (No.17K14270 [NN];
No.16H06490 [YT]).


{\small 
\bibliographystyle{notitle}
\bibliography{references}
}

\end{document}